\documentclass[a4paper,11pt,amsmath,amssymb,superscriptaddress] {revtex4}
\usepackage{graphicx, subfigure}
\usepackage{dcolumn}
\usepackage{color}
\usepackage{bm}
\usepackage[T1]{fontenc}
\usepackage[matrix,frame,arrow]{xypic}
\input{Qcircuit}

\newcommand{\minou}{\text{-}}
\newcommand{\plusou}{\text{+}}

\newcommand{\abs}[1]{\left|#1\right|}

\newcommand{\ket}[1]{\left|#1\right>}
\newcommand{\bra}[1]{\left<#1\right|}

\newcommand{\bpi}{\text{\bf{$\Pi$}}}
\newcommand{\ba}{\text{\bf{a}}}

\newcommand{\bH}{\text{\bf{H}}}
\newcommand{\bSz}{\text{\boldsymbol{$\sigma_z$}}}
\newcommand{\cU}{\mathcal{U}}

\newcommand{\KK}{{\mathcal K}}
\newcommand{\NN}{{\mathcal N}}
\newcommand{\CC}{{\mathcal C}}

\begin{document}

\title{Hardware-efficient autonomous quantum error correction}
\author{Zaki Leghtas}
\affiliation{INRIA Paris-Rocquencourt, Domaine de Voluceau, B.P.~105, 78153 Le Chesnay Cedex, France}
\author{Gerhard Kirchmair}
\affiliation{Department of Physics and Applied Physics, Yale University, New Haven, Connecticut 06520, USA}
\author{Brian Vlastakis}
\affiliation{Department of Physics and Applied Physics, Yale University, New Haven, Connecticut 06520, USA}
\author{Robert Schoelkopf}
\affiliation{Department of Physics and Applied Physics, Yale University, New Haven, Connecticut 06520, USA}
\author{Michel Devoret}
\affiliation{Department of Physics and Applied Physics, Yale University, New Haven, Connecticut 06520, USA}
\author{Mazyar Mirrahimi}
\affiliation{INRIA Paris-Rocquencourt, Domaine de Voluceau, B.P.~105, 78153 Le Chesnay Cedex, France}
\affiliation{Department of Physics and Applied Physics, Yale University, New Haven, Connecticut 06520, USA}

\date{\today}

\begin{abstract}
We propose a new method to autonomously correct for errors of a logical qubit induced by energy relaxation. This scheme encodes the logical qubit  as a multi-component superposition of coherent states in a harmonic oscillator, more specifically a cavity mode. The sequences of encoding, decoding and correction operations employ the non-linearity provided by a single physical qubit coupled to the cavity. We layout in detail how to implement these operations in a practical system. This proposal directly addresses the task of building a hardware-efficient and technically realizable quantum memory.
\end{abstract}

\maketitle

\section{Introduction}
Long lived coherence is a prerequisite for quantum computation. The last two decades have seen impressive improvements in the coherence times of qubits and cavities. The results of hardware improvement have been so substantial that the quality threshold needed for quantum error correction (QEC) \cite{Shor-QEC,Steane-PRL_1996} to be effective is within reach \cite{Paik-et-al-PRL2011,steffen-viewpoint-2011}. Since the birth of QEC, many possible implementations have been proposed. We classify these QEC schemes as being either measurement based or autonomous. Measurement based QEC (MBQEC) consists of periodically measuring error syndromes and feeding back appropriate correction pulses conditioned by the measurement results \cite{Chiaverini-al-Wineland-Nature_2004}. In autonomous QEC (AQEC), however, no classical information needs to be extracted. Instead, it is sufficient to transfer the random errors to an ancillary quantum system which is then reset to remove the entropy~\cite{Schindler-al-Blatt-Science_2011,Reed-et-al-Nature2012}. Furthermore, continuous time implementations can be used for both MBQEC~\cite{Ahn-Doherty-Landahl-PRA_2002} and AQEC~\cite{mabuchi-et-al-PRL2010}.

All these methods encode the single logical qubit to be protected in a register of several physical qubits. While at least five physical qubits are needed to correct for single phase and bit flip errors of the register~\cite{gottesman-96}, it has been shown that when the decoherence is due to a dominant quantum noise process such as amplitude damping, less resources might be needed. Indeed,~\cite{leung-et-al-PRA97} propose a four-qubit code correcting for single amplitude damping errors.

In this paper, we propose a QEC scheme which replaces the register of qubits by a single high-Q cavity mode, coupled to a single physical qubit. The vastness of the Hilbert space of the harmonic oscillator, combined with the control provided by operations described in \cite{Leghtas-al-PRLsubmitted_2012}, allows this replacement. In our scheme, the logical qubit is encoded in a multi-component superposition of coherent states in the cavity mode, the coupled qubit bringing the non-linearity necessary for the manipulation of coherent states. This simple cavity-qubit system is the standard building block of both circuit and cavity quantum electrodynamics (QED) experiments~\cite{Schoelkopf-Girvin-Nature_2008}. Here, we show that this minimal hardware, together with an additional low-Q cavity mode used for qubit readout or qubit reset, is sufficient to correct for the dominant source of errors, namely photon damping in the high-Q cavity. Moreover, the number of independent quantum noise channels corrupting  the logical information, does not increase with the number of encoded qubits, which represents an additional advantage of our protocol over a multi-qubit register. A cavity mode is thus a powerful hardware for protecting quantum information~\cite{gottesman-et-al-01,vitali-et-al-PRA98,zippilli-et-al-03}.

Previous proposals~\cite{vitali-et-al-PRA98,zippilli-et-al-03} have suggested that states with a given photon number parity of a cavity mode (parity states) can be used to encode quantum information. There, stabilization of a parity manifold is obtained by a quantum non-demolition (QND) parity measurement and photon injection when needed. Such a scheme replaces the decoherence due to photon damping by a slower, unusual dephasing due to a drift in the parity manifold.

Here, we go a step forward towards a readily realizable quantum memory, by proposing a scheme for efficiently encoding a logical qubit in a particular cavity state that is fully protected against single quantum jumps due to photon damping. Depending on the experimental constraints, two approaches are possible. The first approach, a MBQEC scheme, keeps track of the quantum jumps by stroboscopic QND measurements of the photon number parity and corrects for the decay by an appropriate decoding and encoding operation sequence. The second approach, an AQEC scheme, transfers the entropy of the cavity state to an ancilla qubit and then removes this entropy by reseting the qubit state. All the encoding, decoding and correction operations of both approaches can be performed using tools that have been introduced in our recent paper \cite{Leghtas-al-PRLsubmitted_2012}.


\section{Cavity logical 1 and logical 0, and MBQEC}
\label{sec:logical0logical1MBQEC}
An arbitrary qubit state $c_g\ket{g}+c_e\ket{e}$ (we denote $\ket{g}$ and $\ket{e}$ the ground and excited state) is mapped into a multi-component coherent state $\ket{\psi_\alpha^{(0)}}=c_g\ket{\CC^+_{\alpha}}+c_e\ket{\CC^+_{i\alpha}}$, where
\begin{equation*}
  \ket{\CC^\pm_{\alpha}} =\NN(\ket{\alpha}\pm\ket{-\alpha})\;,\;\;\;\ket{\CC^\pm_{i\alpha}}=\NN(\ket{i\alpha}\pm\ket{-i\alpha})\;.
\end{equation*}
$\NN$ ($\approx 1/\sqrt 2$) is a normalizing factor and $\ket{\alpha}$ denotes a coherent state of complex amplitude $\alpha$, chosen such that $\ket{\alpha}, \ket{-\alpha}, \ket{i\alpha}, \ket{-i\alpha}$ are quasi-orthogonal. Together with $\ket{\psi_\alpha^{(0)}}$, we introduce
\begin{eqnarray*}
  \ket{\psi_\alpha^{(1)}}&=&c_g\ket{\CC^-_{\alpha}}+ic_e\ket{\CC^-_{i\alpha}}\\
  \ket{\psi_\alpha^{(2)}}&=&c_g\ket{\CC^+_{\alpha}}-c_e\ket{\CC^+_{i\alpha}}\\
  \ket{\psi_\alpha^{(3)}}&=&c_g\ket{\CC^-_{\alpha}}-ic_e\ket{\CC^-_{i\alpha}}\;.
\end{eqnarray*}
We have recently proposed a toolbox of operations which efficiently prepare the states $\ket{\psi_\alpha^{(n)}}$ \cite{Leghtas-al-PRLsubmitted_2012}. The logical 0, $\ket{\CC^+_{\alpha}}$, and the logical 1, $\ket{\CC^+_{i\alpha}}$, have the three following remarkable properties: first, the states $\ket{\psi_\alpha^{(n)}}$ evolve after a quantum jump due to a photon loss, to
\begin{equation*}
  \ba\ket{\psi_\alpha^{(n)}}/\left\|\ba\ket{\psi_\alpha^{(n)}}\right\|=\ket{\psi_\alpha^{\left((n+1) \text{ mod }4\right)}}\;,
\end{equation*}
where $\ba$ is the annihilation operator. Therefore the set $\{\ket{\psi_\alpha^{(n)}}\}$ is closed under the action of $\ba$. Second, in absence of jumps during a time interval $t$, $\ket{\psi_\alpha^{(n)}}$ deterministically evolves to $\ket{\psi_{\alpha e^{-\kappa t/2}}^{(n)}}$, where $\kappa$ is the cavity decay rate. Third, defining the parity operator $\bpi=\exp(i\pi\ba^\dag \ba)$, we have
\begin{equation*}
  \bra{\psi_\alpha^{(n)}}\bpi\ket{\psi_\alpha^{(n)}}=(-1)^n\;.
\end{equation*}
The parity operator acts therefore as a quantum jump indicator. Now, suppose we have a QND parity measurement, and that we have counted $c$ jumps during a time $t$, the initial state has evolved to $\ket{\psi_{\alpha e^{-\kappa t/2}}^{(c \text{ mod } 4)}}$. Using operations introduced in \cite{Leghtas-al-PRLsubmitted_2012}, we can find a unitary transformation, independent of $c_g$ and $c_e$, which maps $\ket{\psi_{\alpha e^{-\kappa t/2}}^{(c \text{ mod } 4)}}$ back to $\ket{\psi_{\alpha}^{(0)}}$, therefore undoing the effect of decoherence. How to find this correcting transformation will be discussed in section \ref{sec:operations}.

This MBQEC we have outlined so far requires a high throughput QND parity measurement and a low latency feedback loop. Let us note that QND parity measurements have been previously performed through Ramsey-type experiments within the context of cavity QED with Rydberg atoms~\cite{haroche-et-al-2007}. Such a measurement scheme can be also adapted to circuit QED experiments but necessitates a flux bias line which would allow one to alternately tune the qubit near or far from resonance with the cavity mode. Also, fast and reliable measurements would necessitate application of quantum limited amplifiers. In the next section, we introduce an AQEC scheme which does not necessitate such resources. Instead, it requires the availability of a rapid, high fidelity qubit reset~\cite{Reed-et-al-APL2010}.

\section{Autonomous QEC}
AQEC is realized by using an auxiliary quantum system that we take here to be the same coupled physical qubit, which is used to manipulate the cavity state. The idea consists in finding a unitary operation $\cU_{\text{correct}}$  such that
\begin{align}
\label{eq:Ucorrect}
\cU_{\text{correct}}:& \ket{g}\otimes \ket{\CC^{\pm}_{\alpha e^{-\kappa t/2}}}\rightarrow  \frac{1}{\sqrt 2}(\ket{g}\pm \ket{e})\otimes \ket{\CC^{+}_\alpha},\\
&  \ket{g}\otimes \ket{\CC^{\pm}_{i\alpha e^{-\kappa t/2}}}\rightarrow  \frac{1}{\sqrt 2}(\ket{g}\pm \ket{e})\otimes \ket{\CC^{+}_{i\alpha}}.\nonumber
\end{align}
This unitary operation transfers the entropy of the quantum system to be protected to the auxiliary one. Now, resetting the state of the auxiliary system, we can evacuate the entropy, restoring the initial full state.

\begin{figure}
  \includegraphics[width=.7\columnwidth]{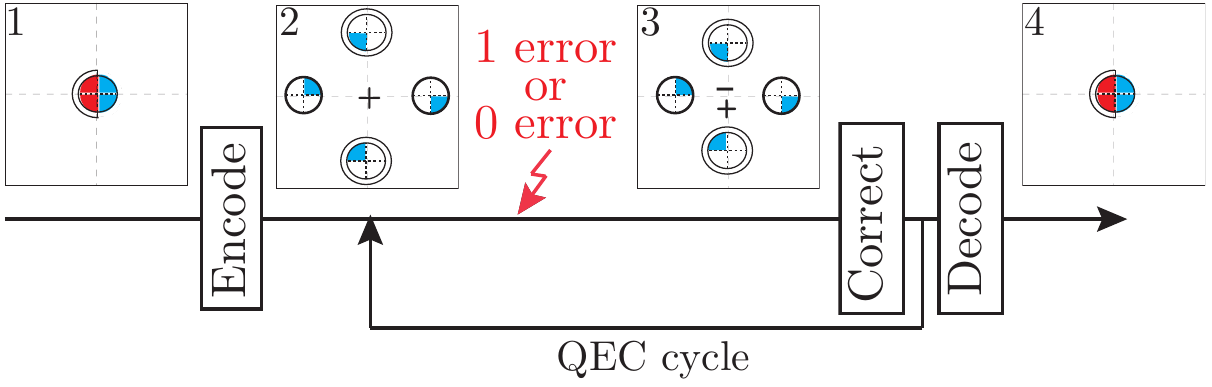}
  \caption{\label{fig:scheme} Our AQEC scheme is composed of three operations: encoding, correcting and decoding. The joint cavity-qubit state is represented by a generalized Fresnel diagram. Fresnel plane positions carry the description of the cavity mode, while colors carry the description of the qubit. Our protocol only requires that we represent superpositions of coherent states entangled with the qubit degrees of freedom. A circle whose center is positioned at $\alpha$ in the diagram corresponds to a coherent state component of amplitude $\alpha$. For example, the diagram in frame 2 represents the state $c_g\ket{C_\alpha^+}+c_e\ket{C_{i\alpha}^+}=\mathcal{N}\left(c_g\ket{g,\alpha}+c_g\ket{g,-\alpha}+c_e\ket{g,i\alpha}+c_e\ket{g,-i\alpha}\right)$, where $\mathcal{N}\approx1/\sqrt{2}$ is a normalization factor. Each component of this state corresponds to a circle whose color refers to whether the qubit is in $\ket{g}$ (blue) or $\ket{e}$ (red). The rim of each circle indicates whether the pre-factor is $c_g$ (single line) or $c_e$ (double line). Finally, the fraction of the colored disc represents the total weight $\left|\mathcal{N}c_{g,e}\right|^2$ of each coherent component. Here, quarter filled circles correspond to $\left|\mathcal{N}c_{g,e}\right|^2=1/4$. Initially (frame 1), the qubit is in $c_g\ket{g}+c_e\ket{e}$ and the cavity is in vacuum. The plus [resp: minus] sign in the 2 and 3 diagrams indicates whether the logical qubit is encoded in the pair $(\ket{C_\alpha^+},\ket{C_{i\alpha}^+})$ [resp: $(\ket{C_\alpha^-},i\ket{C_{i\alpha}^-})$]. A jump from a plus to a minus sign is induced by a photon loss error, which we aim to correct.}
\end{figure}

Here, the AQEC scheme consists of encoding the qubit state $c_g\ket{g}+c_e\ket{e}$ in the state $\ket{\psi_{\alpha}^{(0)}}$ and, in a stroboscopic manner, performing the above unitary transformation followed by the qubit reset. Assuming that at most one quantum jump can happen between two correction operations separated by time $T_w$, the state before the correction is given either by $\ket{\psi_{\alpha e^{-\kappa T_w/2}}^{(0)}}$ or $\ket{\psi_{\alpha e^{-\kappa T_w/2}}^{(1)}}$. After the correction operation, we have restored the initial state $\ket{\psi_\alpha^{(0)}}$. This whole process may be finished by a decoding step transferring the quantum information back onto the qubit (see Fig.~\ref{fig:scheme}).

We now quantify the performance of our AQEC scheme. Let $\rho_{\alpha}^{(n)}$ denote the projector onto the state $\ket{\psi_{\alpha}^{(n)}}$. The effect of the waiting time $T_w$ between two corrections may be modeled by a Kraus operator
\begin{eqnarray*}
  \KK_w: \rho^{(0)}_\alpha&\rightarrow& p_0 \rho^{(0)}_{\tilde\alpha} + p_1 \rho^{(1)}_{\tilde\alpha} +p_{2} \rho^{(2)}_{\tilde\alpha}+p_{3} \rho^{(3)}_{\tilde\alpha}\; ,
\end{eqnarray*}
where $\tilde \alpha=\alpha e^{-\kappa T_w/2}$. For a Poisson process with a jump rate $\lambda_{\text{jump}}$, the probability of having $k$ jumps during a time interval $T_w$ is given by $\exp(-\lambda_{\text{jump}}T_w)\lambda_{\text{jump}}^kT_w^k/k!$. We denote $p_k$ the probability of having $k$ (mod 4) jumps during the waiting time $T_w$. In the limit where $\epsilon_{\text{jump}}\equiv\lambda_{\text{jump}}T_w=\kappa T_w \bar n\ll 1$, we have $p_0\approx 1-\epsilon_{\text{jump}}+\epsilon_{\text{jump}}^2/2$, $p_1\approx \epsilon_{\text{jump}}-\epsilon_{\text{jump}}^2$, $p_2+p_3 \approx \epsilon_{\text{jump}}^2/2$.
The correction step consists of the joint unitary operation on the cavity-qubit system followed by the qubit reset. We model the effect of this operation by the Kraus operator $\KK_c$, mapping both $\ket{\psi_{\alpha e^{-\kappa T_w/2}}^{(0)}}$ and  $\ket{\psi_{\alpha e^{-\kappa T_w/2}}^{(1)}}$ to $\ket{\psi_\alpha^{(0)}}$. After $N$ correction cycles and waiting times (each one taking a time $T_c+T_w$), we obtain a fidelity at time $t_N=N(T_c+T_w)$: $F_\text{AQEC}(t_N)=\abs{\bra{\psi_\alpha^{(0)}} (\KK_c\KK_w)^N \ket{\psi_\alpha^{(0)}}}^2$.

We denote $(1-\epsilon_{\text{correct}})$  the fidelity of the correction operation, taking into account various imperfections and particularly finite coherence times. Also, $\epsilon_{\text{wait}}=\epsilon_{\text{jump}}^2/2$ denotes the probability of having 2 or more jumps during the waiting time between two correction steps. We have $F_\text{AQEC}(t_N)\approx\left((1-\epsilon_{\text{correct}})(1-\epsilon_{\text{wait}})\right)^N$. Assuming $T_c\ll T_w$, we obtain an effective decay rate $\kappa_{eff}\approx (\epsilon_{\text{correct}}+(\kappa T_w \bar n)^2/2)/T_w$. The latter is maximal for $T_w=\sqrt{2\epsilon_{\text{correct}}}/\kappa\bar n$, which would lead to
\begin{equation}
\label{eq:kappaeff}
\kappa_{eff}=\kappa\bar n \sqrt{2\epsilon_{\text{correct}}}\;.
\end{equation}
This is an improvement by a factor of $\sqrt{2\epsilon_{\text{correct}}}$ with respect to the decay rate $\kappa\bar n$ of $\ket{\psi_{\alpha}^{(0)}}$ in absence of correction.

\section{Encoding, decoding and correcting operations}
\label{sec:operations}
\begin{figure*}[ht!]
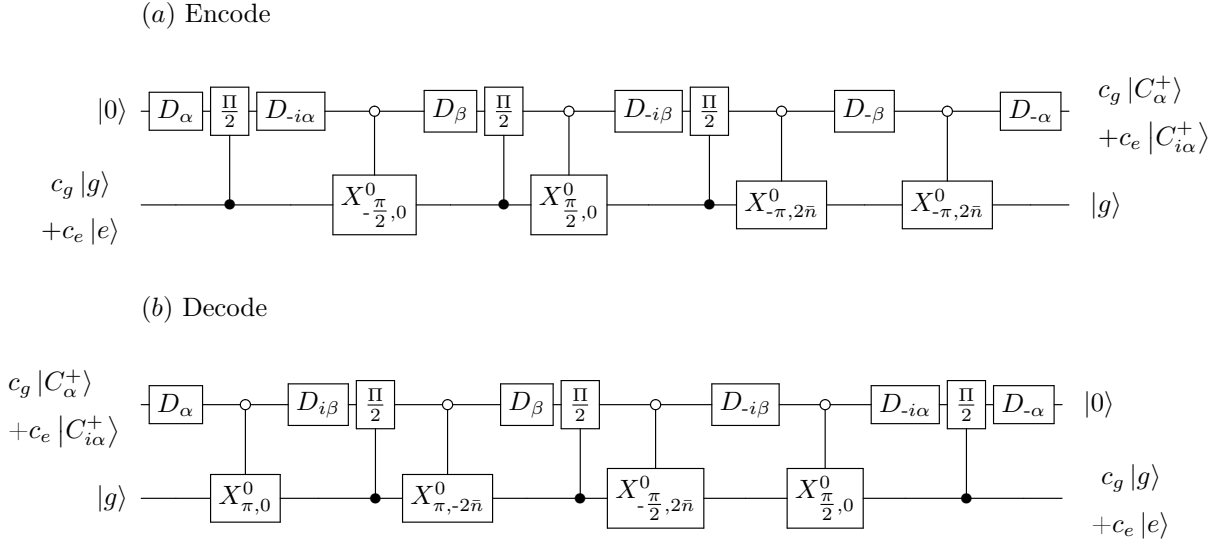

$\begin{array}{lc}
  (a) \text{ Encode}\\
  \; \\
  \input{qubit2cats}\\
  \; \\
  (b) \text{ Decode}\\
  \; \\
  \input{cats2qubit}
\end{array}$
  \caption{Sequence of operations which generate $\cU_{\text{encode}}$ (a) and $\cU_{\text{decode}}$ (b), mapping the qubit state to the cavity and back. $D_\alpha$ displaces the cavity state by an amplitude $\alpha$ regardless of the qubit state. {Conditional} operation $\Pi$ [resp: $\tfrac{\Pi}{2}$] is realized by simply waiting for time $\pi/\chi$ [resp: $\pi/(2\chi)$]. This transforms states of the form $\ket{e,\alpha}$ to $\ket{e,-\alpha}$ [resp $\ket{e,i\alpha}$], and leaves $\ket{g,\alpha}$ unchanged.  The conditional qubit rotation $X^0_{\theta,\eta}$ rotates the qubit state by $e^{\tfrac{\theta}{2}(e^{i\eta}\ket{e}\bra{g}-e^{-i\eta}\ket{g}\bra{e})}$ only if the cavity is in the vacuum state $\ket{0}$. This is achieved by applying a long selective pulse exploiting the energy level dispersive shifts. We denote $\beta=\alpha(-1+i)$ {and $\bar n=|\alpha|^2$}.}
  \label{fig:encodedecode}
\end{figure*}
\begin{figure*}
\setlength{\unitlength}{1cm}
\begin{picture}(15,11)
\put(0,10.8){(a) \text{ Encode}}
\put(0,5.2){(b) \text{ Decode }}
\put(0.5,5.7){\includegraphics[width=15cm]{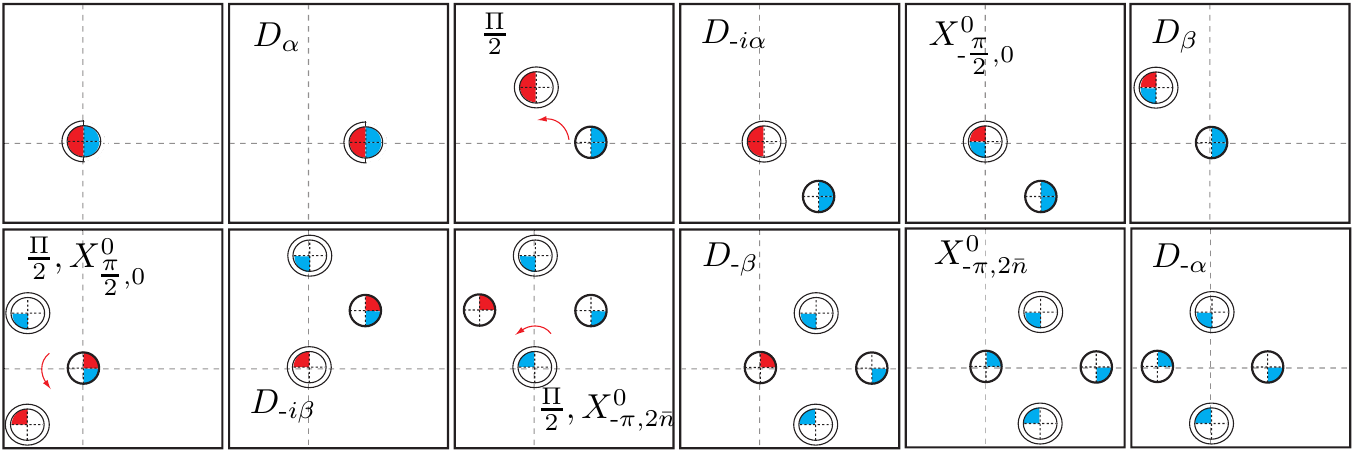}}
\put(0.5,0){\includegraphics[width=15cm]{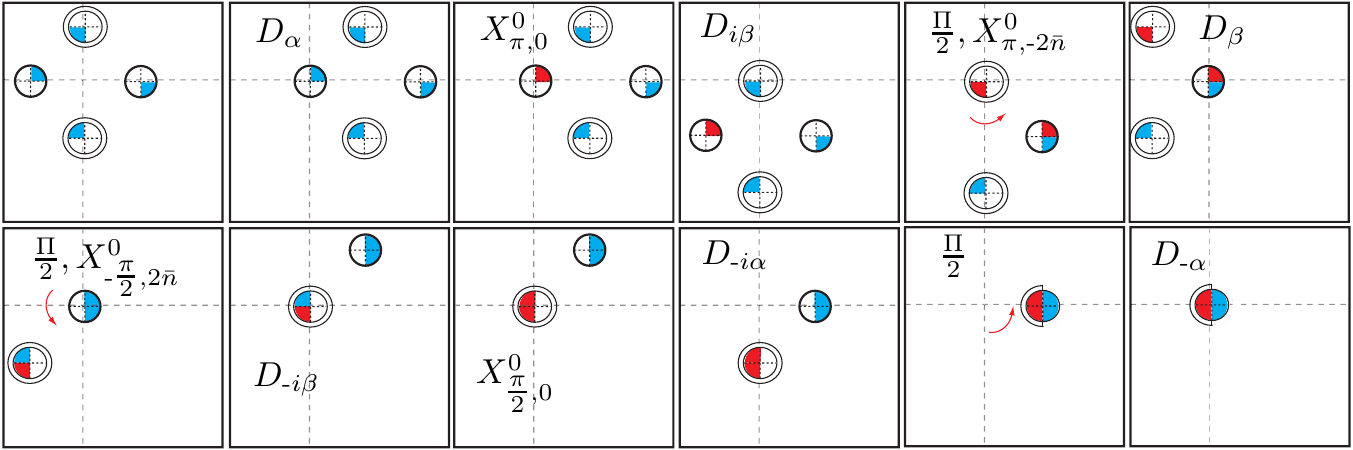}}
\end{picture}
\caption{Diagrammatic equivalent of Fig. \ref{fig:encodedecode}. The frames are ordered from left to right and top to bottom. The diagram notations are explained in Fig. \ref{fig:scheme}. The symbol given in frame $n$ corresponds to the operation performed to go from frame $n-1$ to $n$. All the possible operations are described in the caption of Fig. \ref{fig:encodedecode}. The curved arrow corresponds to the rotation of the excited state component of the state. \label{fig:EncodeDecode_diagrams}}
\end{figure*}
\begin{figure*}[ht!]
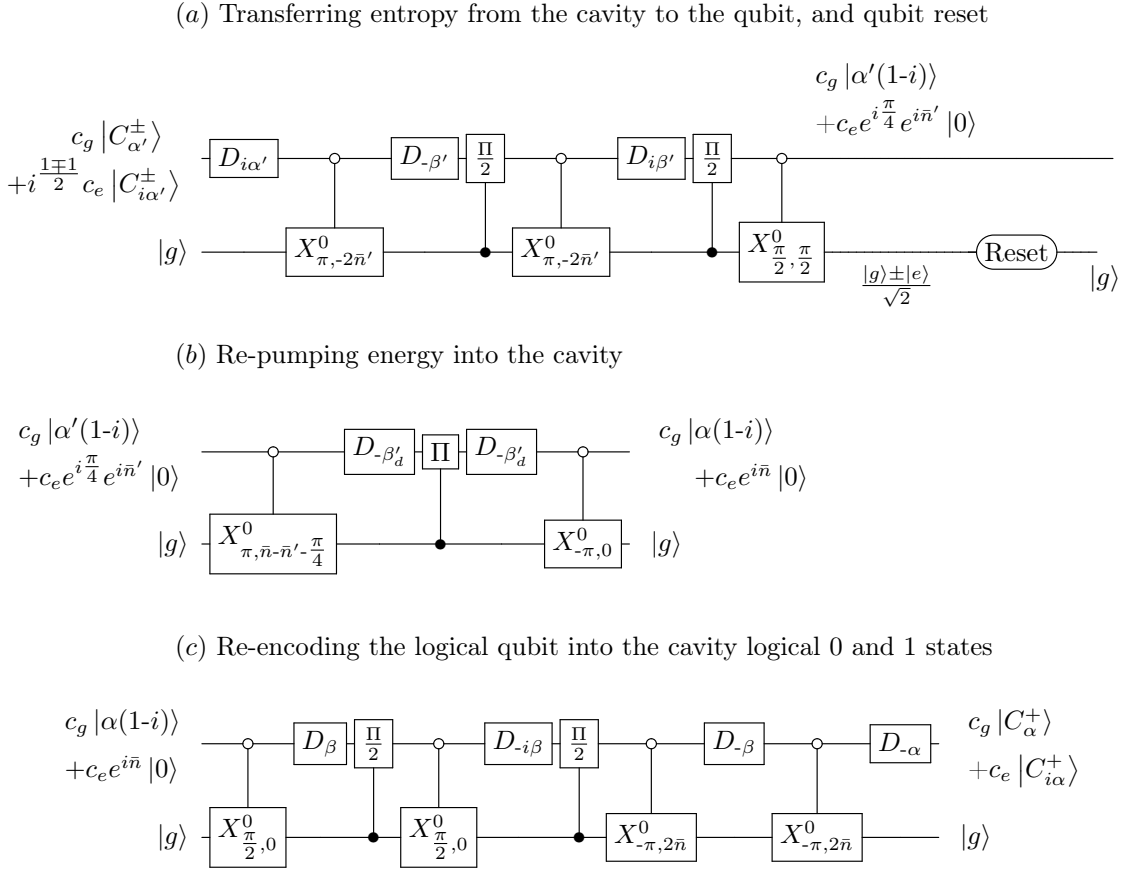

  $\begin{array}{l}
  (a) \text{ Transferring entropy from the cavity to the qubit, and qubit reset} \\
  \;\\
  \;\\
  \quad\input{transferentropy}\\
  \;\\
  (b)\text{ Re-pumping energy into the cavity}\\
  \;\\
  \quad\input{repump}\\
  \;\\
  (c)\text{ Re-encoding the logical qubit into the cavity logical 0 and 1 states}\\
  \;\\
  \quad\input{rebuildcats}\\
  \end{array}$
  \caption{Full correcting sequence obtained by concatenating the three sequences of pulses (a-c). (a) First, the entropy is transferred from the cavity to the qubit, and then, the qubit is reset to its ground state. (b) Energy is re-pumped into the coherent component to compensate the deterministic decay due to damping during the waiting time $T_w$ between two correction sequences. (c) The cavity state is mapped back onto the initial cavity logical 0 and logical 1. See the caption of Fig.~\ref{fig:encodedecode} for a description of operations $D_\alpha,\Pi$ and $X^0_{\theta,\eta}$. Here, we denote $\alpha'=e^{-\kappa T_w/2}\alpha$ the damped amplitude after the waiting time $T_w$, $\bar n'=|\alpha'|^2$ and $\beta'=\alpha'(i-1)$. In order to compensate the damping during $T_w$, during the re-pumping step, we take $\beta'_d=(\beta'-\beta)/2$.}
  \label{fig:Correct}
\end{figure*}

\begin{figure*}[ht!]
\setlength{\unitlength}{1cm}
\begin{picture}(15,15.1)
\put(0,14.3){(a)\text{ Transferring entropy from the cavity to the qubit, and qubit reset}}
\put(0,8.5){(b)\text{ Re-pumping energy into the cavity}}
\put(0,5.3){(c) \text{ Re-encoding the logical qubit into the cavity logical 0 and 1 states}}
\put(0.5,8.9){\includegraphics[width=15cm]{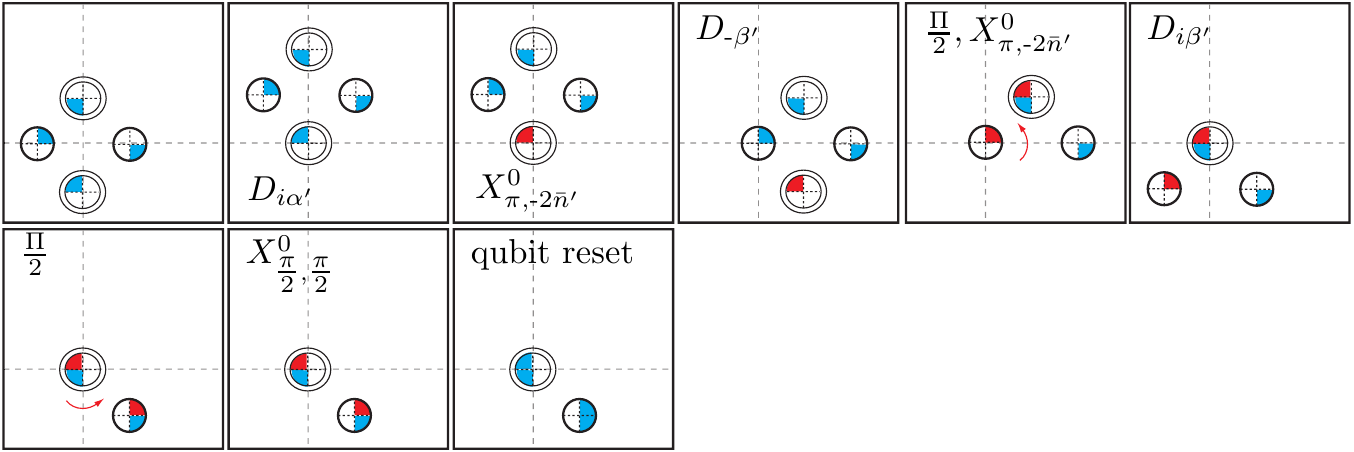}}
\put(0.5,5.7){\includegraphics[width=15cm]{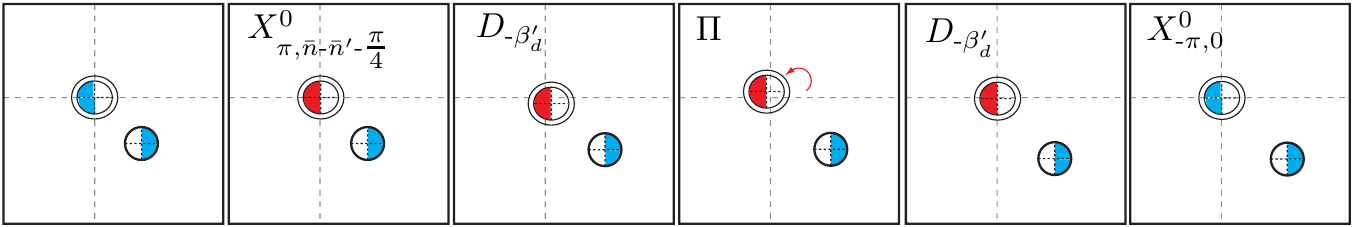}}
\put(0.5,0){\includegraphics[width=15cm]{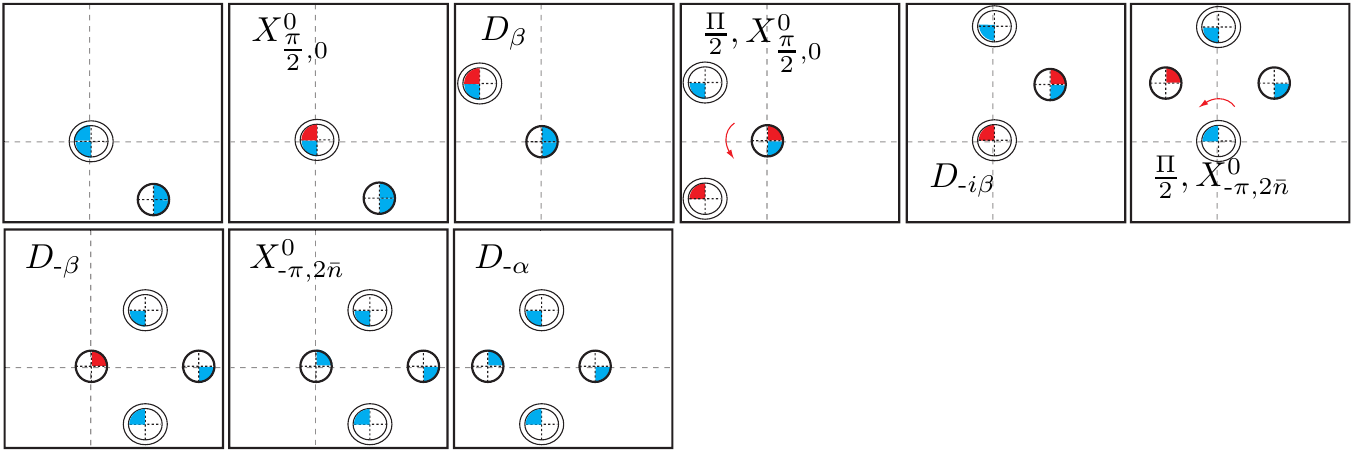}}
\end{picture}
\caption{Diagrammatic equivalent of Fig. \ref{fig:Correct}. See Fig.~\ref{fig:scheme} for an explanation of the diagram notation. In the frame before last of (a), the error is encoded in the phase of the qubit superposition, and is not represented in this diagram. After qubit reset (last of frame of (a)), this phase information is erased.}
\label{fig:Correct_diagram}
\end{figure*}
We define the unitary operations $\cU_{\text{encode}}$ and $\cU_{\text{decode}}$ such that for all $c_g$ and $c_e$,
\begin{align*}
\cU_{\text{\scriptsize encode}}: &(c_g\ket{g}+c_e\ket{e})\otimes \ket{0} \rightarrow \ket{g}\otimes (c_g\ket{\CC^+_{\alpha}}+c_e\ket{\CC^+_{i\alpha}})\;.\\
\cU_{\text{\scriptsize decode}}: &\ket{g}\otimes (c_g\ket{\CC^+_{\alpha}}+c_e\ket{\CC^+_{i\alpha}}) \rightarrow (c_g\ket{g}+c_e\ket{e})\otimes \ket{0}.
\end{align*}

In this section, we show how we could perform $\cU_{\text{\scriptsize encode}}$, $\cU_{\text{\scriptsize decode}}$ and the correcting operation \eqref{eq:Ucorrect}, in practice. We place ourselves in the strong dispersive regime, where both the qubit and the resonator transition frequencies split into well-resolved spectral lines indexed by the number of excitations in the qubit and the resonator~\cite{schuster-nature07}. The resonator frequency $\omega_r$ splits into two well resolved lines $\omega_r^g$ and $\omega_r^e$, corresponding to the cavity's frequency when the qubit is in the ground ($\ket{g}$) or the excited ($\ket{e}$) state. Through the same mechanism, the qubit frequency $\omega_q$ splits into $\{\omega_q^n\}_{n=0,1,2,\cdots}$ corresponding to the qubit frequency when the cavity is in the photon number state $\ket{n}$. Recent experiments have shown dispersive shifts that are more than 3 orders of magnitude larger than the qubit and cavity linewidths~\cite{Paik-et-al-PRL2011}.

The Hamiltonian of such a dispersively coupled qubit-cavity system is well approximated by
\begin{equation*}
  \bH_0=\omega_q \frac{\bSz}{2}+\omega_c \ba^\dag\ba - \chi \frac{\bSz}{2}\ba^\dag\ba\;,
\end{equation*}
where $\omega_q$ and $\omega_c$ are respectively the qubit and cavity frequencies, $\chi$ is the dispersive coupling and $\bSz=\ket{e}\bra{e}-\ket{g}\bra{g}$. This Hamiltonian may be written in an appropriate rotating frame as $\bH=-\chi \ket{e}\bra{e}\ba^\dag\ba$. This dispersive coupling is called strong when $\chi\gg \kappa,1/T_2$, where $T_2=(1/2T_1+1/T_{\phi})^{-1}$ is the qubit decoherence time.

{As detailed in~\cite{Leghtas-al-PRLsubmitted_2012}, the strong dispersive cavity-qubit coupling allows to efficiently perform conditional cavity displacements and conditional qubit rotations. Long selective qubit pulses with carrier frequency $\omega_q^0$ can rotate the qubit state conditioned on the cavity being in the vacuum state. Similarly, selective cavity pulses with carrier frequency $\omega_r^g$ [resp: $\omega_r^e$] can coherently displace the cavity state conditioned to the qubit being in the ground [resp: excited] state. Furthermore, as explained in~\cite{Caves-Shaji-2010,Leghtas-al-PRLsubmitted_2012}, shorter operation times are obtained by replacing a conditional cavity displacement by two unconditional ones separated by a waiting time.}

The operations involved in our QEC scheme rely on a qubit reset and three unitary transformations. The first one, $D_\alpha$, displaces the cavity state by a {complex} amplitude $\alpha$ regardless of the qubit state. Second, the conditional operation $\Pi$ [resp: $\tfrac{\Pi}{2}$] transforms states of the form $\ket{e,\alpha}$ to $\ket{e,-\alpha}$ [resp $\ket{e,i\alpha}$], and leaves $\ket{g,\alpha}$ unchanged. It is realized by simply waiting for time $\pi/\chi$ [resp: $\pi/(2\chi)$]. Third, a conditional qubit rotation $X^0_{\theta,\eta}$ rotates the qubit state by $e^{\tfrac{\theta}{2}(e^{i\eta}\ket{e}\bra{g}-e^{-i\eta}\ket{g}\bra{e})}$ only if the cavity is in the vacuum state $\ket{0}$. This is achieved by applying a long selective pulse exploiting the energy level dispersive shifts.

The reset operation forces the qubit state to $\ket{g}$ independently of the cavity state. This operation needs to be fast compared to $\chi$ to avoid re-entanglement of the qubit to the cavity mode. A possible scheme to perform such a fast reset is to rapidly tune (e.g. with a flux bias line) the qubit frequency to bring it into resonance with a low-Q cavity mode~\cite{Reed-et-al-APL2010}. Another possibility, avoiding fast frequency tuning, is to perform a dynamical cooling cycle as proposed in~\cite{Leghtas-et-al-APS-2012}. See Fig.~\ref{fig:encodedecode} and Fig.~\ref{fig:Correct} for a detailed illustration of how combining all these operations leads to the encoding, decoding and correcting gates. A graphical representation of the sequence of operations is given in Fig.~\ref{fig:EncodeDecode_diagrams} and Fig.~\ref{fig:Correct_diagram}. In the correction sequence (see Figs.~\ref{fig:Correct},~\ref{fig:Correct_diagram}), we have introduced the qubit reset in the middle, in contrast with what Eq.~\eqref{eq:Ucorrect} suggests. We find the resulting sequence to be more efficient.

Simply adapting the sequence of Fig.~\ref{fig:Correct}, without the qubit reset, we would obtain the correcting sequence suggested in the MBQEC scheme of section \ref{sec:logical0logical1MBQEC}.

\section{Simulations}
\begin{figure}[h]
  \includegraphics[width=.7\columnwidth]{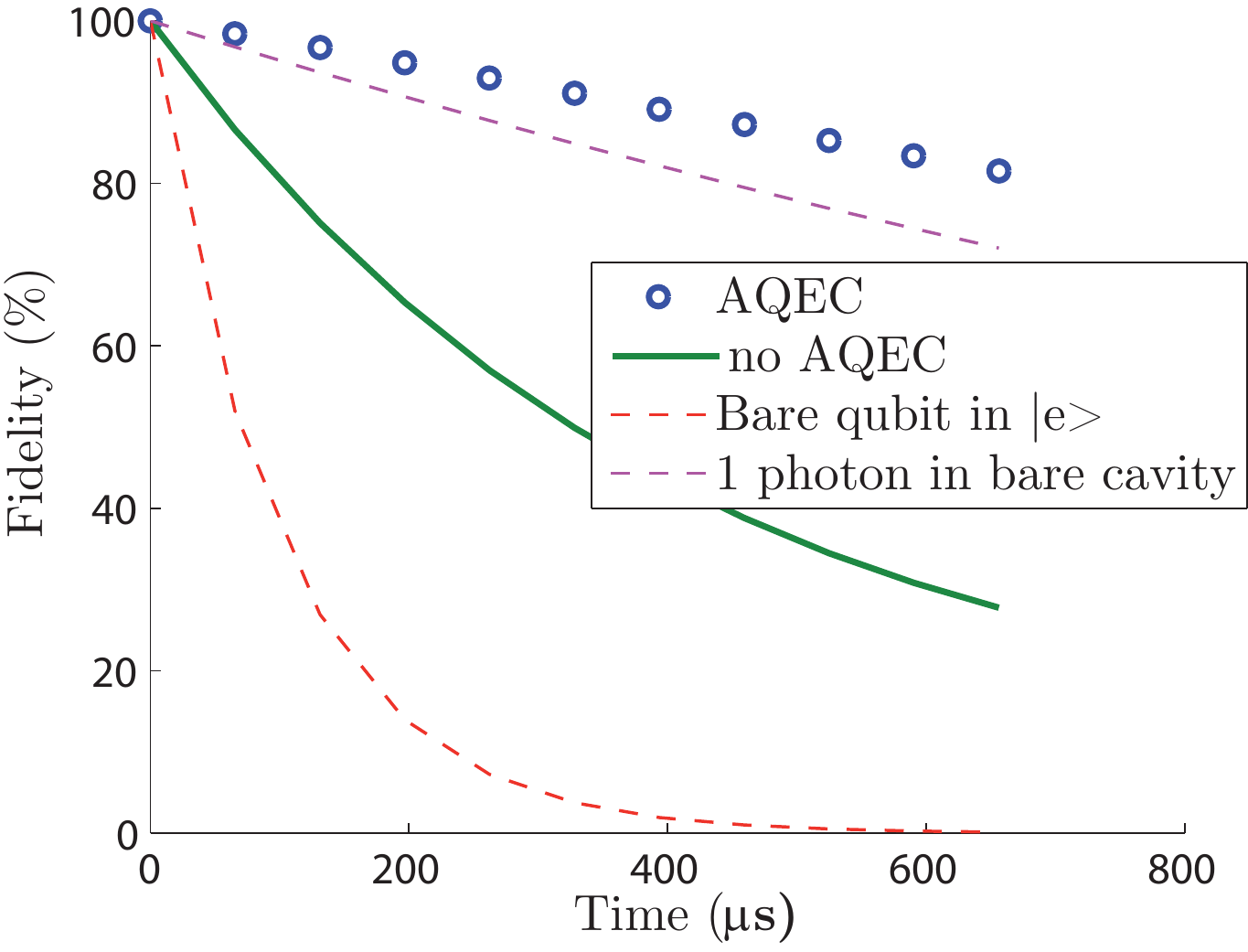}
  \caption{Fidelity of our AQEC scheme. The cavity state is initialized in state $\ket{\psi_\alpha^{(0)}}$.
  We represent the fidelity of the cavity state w.r.t $\ket{\psi_\alpha^{(0)}}$ in absence of the error correction sequence of Fig.~\ref{fig:Correct} (green solid line) and after each correction sequence (blue open dots).
  These simulations take into account dominant sources of decoherence: cavity decay $T_{cav}=2$ ms and qubit $T_1=T_2=100~\mu$s.
   The decay rate of the bare qubit and cavity are plotted for comparison (red and magenta dashed lines).}
  \label{fig:QECFidelities}
\end{figure}
To numerically compute the fidelity of each operation, we simulate the corresponding sequence of pulses on a master equation taking into account the qubit and cavity decoherence. We take a qubit decay time $T_1$ and pure dephasing time $T_2$ of $T_1=T_2=100~\mu$s and a cavity lifetime $T_{\text{cav}}=1/\kappa=2~$ms. The dispersive coupling strength is $\chi/2\pi=40$ MHz and the mean photon number per coherent component is $\bar n=|\alpha|^2=4$. We get an encoding and decoding fidelity of $1-\epsilon_{\text{encode}}=1-\epsilon_{\text{decode}}=99.65\%$ for an operation time of $T_e=T_d=231$ns, and a correcting fidelity of $1-\epsilon_{\text{correct}}=99.23\%$ for an operation time of $T_c=519~$ns. The optimal waiting time is $T_w=65.6~\mu$s. This leads to an effective lifetime for the corrected state of $T^{eff}_{cav}=4.1$ ms, within $1\%$ agreement with the formula in \eqref{eq:kappaeff}. This represents a large improvement over the lifetime of the bare qubit $T_1=100~\mu$s, and of an uncorrected cavity $T_{cav}/\bar n=500~\mu$s. There is even an improvement w.r.t the single photon lifetime of $T_{cav}=2$ ms. Note that a better correction gate fidelity could lead to an even larger improvement. This could be obtained by optimizing the pulse sequence for a specific experimental setting.

\section{Conclusion}
We have shown that it is possible to protect a logical qubit against relaxation by encoding it in a single cavity coupled to a single physical qubit, and driving them with simple control pulses.
As long as the cavity-qubit coupling is in the strong dispersive regime, no control over this coupling is necessary and the autonomous error correction may be performed without any real-time qubit frequency tuning.
Our theoretical prediction of the lifetime improvement is confirmed by numerical simulations of the proposed protocol.
Also, additional control of the qubit frequency in real time could lead to simpler and faster operations with higher fidelities, by using ideas similar to those in \cite{Caves-Shaji-2010}.
Finally, our scheme that corrects only for single jumps in the cavity, could be generalized to an $n^{th}$ order correcting scheme by superposing $n+1$ quasi-orthogonal coherent states for each logical state zero and one.

\textbf{Acknowledgements}
{This work was partially supported by the French ``Agence Nationale de la Recherche'' under the project EPOQ2 number ANR-09-JCJC-0070 and
the Army Research Office (ARO) under the project number ARO - W911NF-09-1-0514. B.V acknowledges support from the National Science Foundation, project number: PHY-0969725.}


\begin{thebibliography}{21}
\expandafter\ifx\csname natexlab\endcsname\relax\def\natexlab#1{#1}\fi
\expandafter\ifx\csname bibnamefont\endcsname\relax
  \def\bibnamefont#1{#1}\fi
\expandafter\ifx\csname bibfnamefont\endcsname\relax
  \def\bibfnamefont#1{#1}\fi
\expandafter\ifx\csname citenamefont\endcsname\relax
  \def\citenamefont#1{#1}\fi
\expandafter\ifx\csname url\endcsname\relax
  \def\url#1{\texttt{#1}}\fi
\expandafter\ifx\csname urlprefix\endcsname\relax\def\urlprefix{URL }\fi
\providecommand{\bibinfo}[2]{#2}
\providecommand{\eprint}[2][]{\url{#2}}

\bibitem[{\citenamefont{Shor}(1995)}]{Shor-QEC}
\bibinfo{author}{\bibfnamefont{P.}~\bibnamefont{Shor}}, \bibinfo{journal}{Phys.
  Rev. A} \textbf{\bibinfo{volume}{52}}, \bibinfo{pages}{2493}
  (\bibinfo{year}{1995}).

\bibitem[{\citenamefont{Steane}(1996)}]{Steane-PRL_1996}
\bibinfo{author}{\bibfnamefont{A.}~\bibnamefont{Steane}},
  \bibinfo{journal}{Phys. Rev. Lett} \textbf{\bibinfo{volume}{77}}
  (\bibinfo{year}{1996}).

\bibitem[{\citenamefont{Paik et~al.}(2011)\citenamefont{Paik, Schuster, Bishop,
  Kirchmair, Catelani, Sears, Johnson, Reagor, Frunzio, Glazman
  et~al.}}]{Paik-et-al-PRL2011}
\bibinfo{author}{\bibfnamefont{H.}~\bibnamefont{Paik}},
  \bibinfo{author}{\bibfnamefont{D.}~\bibnamefont{Schuster}},
  \bibinfo{author}{\bibfnamefont{L.}~\bibnamefont{Bishop}},
  \bibinfo{author}{\bibfnamefont{G.}~\bibnamefont{Kirchmair}},
  \bibinfo{author}{\bibfnamefont{G.}~\bibnamefont{Catelani}},
  \bibinfo{author}{\bibfnamefont{A.}~\bibnamefont{Sears}},
  \bibinfo{author}{\bibfnamefont{B.}~\bibnamefont{Johnson}},
  \bibinfo{author}{\bibfnamefont{M.}~\bibnamefont{Reagor}},
  \bibinfo{author}{\bibfnamefont{L.}~\bibnamefont{Frunzio}},
  \bibinfo{author}{\bibfnamefont{L.}~\bibnamefont{Glazman}},
  \bibnamefont{et~al.}, \bibinfo{journal}{Phys. Rev. Lett.}
  \textbf{\bibinfo{volume}{107}}, \bibinfo{pages}{240501}
  (\bibinfo{year}{2011}).

\bibitem[{\citenamefont{Steffen}(2011)}]{steffen-viewpoint-2011}
\bibinfo{author}{\bibfnamefont{M.}~\bibnamefont{Steffen}},
  \bibinfo{journal}{Physics} \textbf{\bibinfo{volume}{4}}, \bibinfo{pages}{103}
  (\bibinfo{year}{2011}).

\bibitem[{\citenamefont{Chiaverini et~al.}(2004)\citenamefont{Chiaverini,
  Leibfried, Schaetz, Barrett, Blakestad, Britton, Itano, Jost, Knill, Langer
  et~al.}}]{Chiaverini-al-Wineland-Nature_2004}
\bibinfo{author}{\bibfnamefont{J.}~\bibnamefont{Chiaverini}},
  \bibinfo{author}{\bibfnamefont{D.}~\bibnamefont{Leibfried}},
  \bibinfo{author}{\bibfnamefont{T.}~\bibnamefont{Schaetz}},
  \bibinfo{author}{\bibfnamefont{M.}~\bibnamefont{Barrett}},
  \bibinfo{author}{\bibfnamefont{R.}~\bibnamefont{Blakestad}},
  \bibinfo{author}{\bibfnamefont{J.}~\bibnamefont{Britton}},
  \bibinfo{author}{\bibfnamefont{W.}~\bibnamefont{Itano}},
  \bibinfo{author}{\bibfnamefont{J.}~\bibnamefont{Jost}},
  \bibinfo{author}{\bibfnamefont{E.}~\bibnamefont{Knill}},
  \bibinfo{author}{\bibfnamefont{C.}~\bibnamefont{Langer}},
  \bibnamefont{et~al.}, \bibinfo{journal}{Nature}
  \textbf{\bibinfo{volume}{432}}, \bibinfo{pages}{602} (\bibinfo{year}{2004}).

\bibitem[{\citenamefont{Schindler et~al.}(2011)\citenamefont{Schindler,
  Barreiro, Monz, Nebendahl, Nigg, Chwalla, Hennrich, and
  Blatt}}]{Schindler-al-Blatt-Science_2011}
\bibinfo{author}{\bibfnamefont{P.}~\bibnamefont{Schindler}},
  \bibinfo{author}{\bibfnamefont{J.~T.} \bibnamefont{Barreiro}},
  \bibinfo{author}{\bibfnamefont{T.}~\bibnamefont{Monz}},
  \bibinfo{author}{\bibfnamefont{V.}~\bibnamefont{Nebendahl}},
  \bibinfo{author}{\bibfnamefont{D.}~\bibnamefont{Nigg}},
  \bibinfo{author}{\bibfnamefont{M.}~\bibnamefont{Chwalla}},
  \bibinfo{author}{\bibfnamefont{M.}~\bibnamefont{Hennrich}}, \bibnamefont{and}
  \bibinfo{author}{\bibfnamefont{R.}~\bibnamefont{Blatt}},
  \bibinfo{journal}{Science} \textbf{\bibinfo{volume}{332}}
  (\bibinfo{year}{2011}).

\bibitem[{\citenamefont{Reed et~al.}(2012)\citenamefont{Reed, DiCarlo, Nigg,
  Sun, Frunzio, Girvin, and Schoelkopf}}]{Reed-et-al-Nature2012}
\bibinfo{author}{\bibfnamefont{M.}~\bibnamefont{Reed}},
  \bibinfo{author}{\bibfnamefont{L.}~\bibnamefont{DiCarlo}},
  \bibinfo{author}{\bibfnamefont{S.}~\bibnamefont{Nigg}},
  \bibinfo{author}{\bibfnamefont{L.}~\bibnamefont{Sun}},
  \bibinfo{author}{\bibfnamefont{L.}~\bibnamefont{Frunzio}},
  \bibinfo{author}{\bibfnamefont{S.}~\bibnamefont{Girvin}}, \bibnamefont{and}
  \bibinfo{author}{\bibfnamefont{R.}~\bibnamefont{Schoelkopf}},
  \bibinfo{journal}{Nature} \textbf{\bibinfo{volume}{482}},
  \bibinfo{pages}{382} (\bibinfo{year}{2012}).

\bibitem[{\citenamefont{Ahn et~al.}(2002)\citenamefont{Ahn, Doherty, and
  Landahl}}]{Ahn-Doherty-Landahl-PRA_2002}
\bibinfo{author}{\bibfnamefont{C.}~\bibnamefont{Ahn}},
  \bibinfo{author}{\bibfnamefont{A.~C.} \bibnamefont{Doherty}},
  \bibnamefont{and} \bibinfo{author}{\bibfnamefont{A.~J.}
  \bibnamefont{Landahl}}, \bibinfo{journal}{Phys. Rev. A.}
  \textbf{\bibinfo{volume}{65}} (\bibinfo{year}{2002}).

\bibitem[{\citenamefont{Kerckhoff et~al.}(2010)\citenamefont{Kerckhoff, Nurdin,
  Pavlichin, and Mabuchi}}]{mabuchi-et-al-PRL2010}
\bibinfo{author}{\bibfnamefont{J.}~\bibnamefont{Kerckhoff}},
  \bibinfo{author}{\bibfnamefont{H.}~\bibnamefont{Nurdin}},
  \bibinfo{author}{\bibfnamefont{D.}~\bibnamefont{Pavlichin}},
  \bibnamefont{and} \bibinfo{author}{\bibfnamefont{H.}~\bibnamefont{Mabuchi}},
  \bibinfo{journal}{Physical Review Letters} \textbf{\bibinfo{volume}{105}},
  \bibinfo{pages}{040502} (\bibinfo{year}{2010}).

\bibitem[{\citenamefont{Gottesman}(1996)}]{gottesman-96}
\bibinfo{author}{\bibfnamefont{D.}~\bibnamefont{Gottesman}},
  \bibinfo{journal}{Phys. Rev. A} \textbf{\bibinfo{volume}{54}},
  \bibinfo{pages}{1862} (\bibinfo{year}{1996}).

\bibitem[{\citenamefont{Leung et~al.}(1997)\citenamefont{Leung, Nielsen,
  Chuang, and Yamamoto}}]{leung-et-al-PRA97}
\bibinfo{author}{\bibfnamefont{D.}~\bibnamefont{Leung}},
  \bibinfo{author}{\bibfnamefont{M.}~\bibnamefont{Nielsen}},
  \bibinfo{author}{\bibfnamefont{I.}~\bibnamefont{Chuang}}, \bibnamefont{and}
  \bibinfo{author}{\bibfnamefont{Y.}~\bibnamefont{Yamamoto}},
  \bibinfo{journal}{Phys. Rev. A} \textbf{\bibinfo{volume}{56}},
  \bibinfo{pages}{2567} (\bibinfo{year}{1997}).

\bibitem[{\citenamefont{Leghtas
  et~al.}(2012{\natexlab{a}})\citenamefont{Leghtas, Kirchmair, Vlastakis,
  Devoret, Schoelkopf, and Mirrahimi}}]{Leghtas-al-PRLsubmitted_2012}
\bibinfo{author}{\bibfnamefont{Z.}~\bibnamefont{Leghtas}},
  \bibinfo{author}{\bibfnamefont{G.}~\bibnamefont{Kirchmair}},
  \bibinfo{author}{\bibfnamefont{B.}~\bibnamefont{Vlastakis}},
  \bibinfo{author}{\bibfnamefont{M.}~\bibnamefont{Devoret}},
  \bibinfo{author}{\bibfnamefont{R.}~\bibnamefont{Schoelkopf}},
  \bibnamefont{and}
  \bibinfo{author}{\bibfnamefont{M.}~\bibnamefont{Mirrahimi}},
  \bibinfo{journal}{Submitted, ar{X}iv:1205.2401}
  (\bibinfo{year}{2012}{\natexlab{a}}).

\bibitem[{\citenamefont{Schoelkopf and
  Girvin}(2008)}]{Schoelkopf-Girvin-Nature_2008}
\bibinfo{author}{\bibfnamefont{R.~J.} \bibnamefont{Schoelkopf}}
  \bibnamefont{and} \bibinfo{author}{\bibfnamefont{S.}~\bibnamefont{Girvin}},
  \bibinfo{journal}{Nature} \textbf{\bibinfo{volume}{451}},
  \bibinfo{pages}{664} (\bibinfo{year}{2008}).

\bibitem[{\citenamefont{Gottesman et~al.}(2001)\citenamefont{Gottesman, Kitaev,
  and Preskill}}]{gottesman-et-al-01}
\bibinfo{author}{\bibfnamefont{D.}~\bibnamefont{Gottesman}},
  \bibinfo{author}{\bibfnamefont{A.}~\bibnamefont{Kitaev}}, \bibnamefont{and}
  \bibinfo{author}{\bibfnamefont{J.}~\bibnamefont{Preskill}},
  \bibinfo{journal}{Phys. Rev. A} \textbf{\bibinfo{volume}{64}},
  \bibinfo{pages}{012310} (\bibinfo{year}{2001}).

\bibitem[{\citenamefont{Vitali et~al.}(1998)\citenamefont{Vitali, Tombesi, and
  Milburn}}]{vitali-et-al-PRA98}
\bibinfo{author}{\bibfnamefont{D.}~\bibnamefont{Vitali}},
  \bibinfo{author}{\bibfnamefont{P.}~\bibnamefont{Tombesi}}, \bibnamefont{and}
  \bibinfo{author}{\bibfnamefont{G.}~\bibnamefont{Milburn}},
  \bibinfo{journal}{Phys. Rev. A} \textbf{\bibinfo{volume}{57}},
  \bibinfo{pages}{4930} (\bibinfo{year}{1998}).

\bibitem[{\citenamefont{Zippilli et~al.}(2003)\citenamefont{Zippilli, Vitali,
  Tombesi, and Raimond}}]{zippilli-et-al-03}
\bibinfo{author}{\bibfnamefont{S.}~\bibnamefont{Zippilli}},
  \bibinfo{author}{\bibfnamefont{D.}~\bibnamefont{Vitali}},
  \bibinfo{author}{\bibfnamefont{P.}~\bibnamefont{Tombesi}}, \bibnamefont{and}
  \bibinfo{author}{\bibfnamefont{J.}~\bibnamefont{Raimond}},
  \bibinfo{journal}{Phys. Rev. A} \textbf{\bibinfo{volume}{67}},
  \bibinfo{pages}{052101} (\bibinfo{year}{2003}).

\bibitem[{\citenamefont{Haroche et~al.}(2007)\citenamefont{Haroche, Brune, and
  Raimond}}]{haroche-et-al-2007}
\bibinfo{author}{\bibfnamefont{S.}~\bibnamefont{Haroche}},
  \bibinfo{author}{\bibfnamefont{M.}~\bibnamefont{Brune}}, \bibnamefont{and}
  \bibinfo{author}{\bibfnamefont{J.}~\bibnamefont{Raimond}},
  \bibinfo{journal}{Journal of Modern Optics} \textbf{\bibinfo{volume}{54}},
  \bibinfo{pages}{2101} (\bibinfo{year}{2007}).

\bibitem[{\citenamefont{Reed et~al.}(2010)\citenamefont{Reed, Johnson, Houck,
  DiCarlo, Chow, Schuster, Frunzio, and Schoelkopf}}]{Reed-et-al-APL2010}
\bibinfo{author}{\bibfnamefont{M.}~\bibnamefont{Reed}},
  \bibinfo{author}{\bibfnamefont{B.}~\bibnamefont{Johnson}},
  \bibinfo{author}{\bibfnamefont{A.}~\bibnamefont{Houck}},
  \bibinfo{author}{\bibfnamefont{L.}~\bibnamefont{DiCarlo}},
  \bibinfo{author}{\bibfnamefont{J.}~\bibnamefont{Chow}},
  \bibinfo{author}{\bibfnamefont{D.}~\bibnamefont{Schuster}},
  \bibinfo{author}{\bibfnamefont{L.}~\bibnamefont{Frunzio}}, \bibnamefont{and}
  \bibinfo{author}{\bibfnamefont{R.}~\bibnamefont{Schoelkopf}},
  \bibinfo{journal}{Applied Physics Letters} \textbf{\bibinfo{volume}{96}},
  \bibinfo{pages}{203110} (\bibinfo{year}{2010}).

\bibitem[{\citenamefont{Schuster et~al.}(2007)\citenamefont{Schuster, Houck,
  Schreier, Wallraff, Gambetta, Blais, Frunzio, Majer, Johnson, Devoret
  et~al.}}]{schuster-nature07}
\bibinfo{author}{\bibfnamefont{D.}~\bibnamefont{Schuster}},
  \bibinfo{author}{\bibfnamefont{A.}~\bibnamefont{Houck}},
  \bibinfo{author}{\bibfnamefont{J.}~\bibnamefont{Schreier}},
  \bibinfo{author}{\bibfnamefont{A.}~\bibnamefont{Wallraff}},
  \bibinfo{author}{\bibfnamefont{J.}~\bibnamefont{Gambetta}},
  \bibinfo{author}{\bibfnamefont{A.}~\bibnamefont{Blais}},
  \bibinfo{author}{\bibfnamefont{L.}~\bibnamefont{Frunzio}},
  \bibinfo{author}{\bibfnamefont{J.}~\bibnamefont{Majer}},
  \bibinfo{author}{\bibfnamefont{B.}~\bibnamefont{Johnson}},
  \bibinfo{author}{\bibfnamefont{M.}~\bibnamefont{Devoret}},
  \bibnamefont{et~al.}, \bibinfo{journal}{Nature}
  \textbf{\bibinfo{volume}{445}}, \bibinfo{pages}{515} (\bibinfo{year}{2007}).

\bibitem[{\citenamefont{Caves and Shaji}(2010)}]{Caves-Shaji-2010}
\bibinfo{author}{\bibfnamefont{C.}~\bibnamefont{Caves}} \bibnamefont{and}
  \bibinfo{author}{\bibfnamefont{A.}~\bibnamefont{Shaji}},
  \bibinfo{journal}{Optics Communications} \textbf{\bibinfo{volume}{283}},
  \bibinfo{pages}{695} (\bibinfo{year}{2010}).

\bibitem[{\citenamefont{Leghtas
  et~al.}(2012{\natexlab{b}})\citenamefont{Leghtas, Geerlings, Shankar,
  Mirrahimi, and Devoret}}]{Leghtas-et-al-APS-2012}
\bibinfo{author}{\bibfnamefont{Z.}~\bibnamefont{Leghtas}},
  \bibinfo{author}{\bibfnamefont{K.}~\bibnamefont{Geerlings}},
  \bibinfo{author}{\bibfnamefont{S.}~\bibnamefont{Shankar}},
  \bibinfo{author}{\bibfnamefont{M.}~\bibnamefont{Mirrahimi}},
  \bibnamefont{and} \bibinfo{author}{\bibfnamefont{M.}~\bibnamefont{Devoret}},
  in \emph{\bibinfo{booktitle}{APS March Meeting}}
  (\bibinfo{publisher}{American Physical Society},
  \bibinfo{year}{2012}{\natexlab{b}}).

\end{thebibliography}

\end{document}